# Dynamics of Choline Chloride based Deep Eutectic Solvents: Neutron Scattering Study

Rinesh T.[1,2], H. Srinivasan[1,2] V.K. Sharma[1,2] and S. Mitra[1,2]

[1]*Solid State Physics Division, Bhabha Atomic Research Centre, Trombay, Mumbai, India – 400085*
[2]*Homi Bhabha National Institute, Anushaktinagar, Mumbai 400094, India*
[a)] Corresponding author: smitra@barc.gov.in

**Abstract.** In this study, we investigate the microscopic diffusion dynamics of choline chloride (ChCl) based deep eutectic solvents (DESs) to elucidate the influence of hydrogen bond donor (HBD) identity on the mobility of cholinium ions. The DES systems examined include ethaline, glyceline, and reline, comprising ChCl mixed with ethylene glycol, glycerol, and urea, respectively, in a 1:2 molar ratio. Quasielastic neutron scattering experiments was used to probe the self-diffusion of cholinium ions at molecular length and time scales. The dynamics were modelled as a combination of jump diffusion of the molecular center of mass and localized translation within transient hydrogen-bond cages. Among the three systems, ethaline consistently exhibited the highest cholinium self-diffusion coefficients across all investigated temperatures, attributed to shorter residence times and more frequent molecular jumps. In contrast, reline displayed longer residence times with significantly larger jump length, leading to a temperature-dependent dynamical crossover: While reline and glyceline exhibited comparable diffusivities at low temperatures, reline surpassed glyceline above 330 K. These findings highlight the crucial role of HBD identity in modulating microscopic diffusion and provide valuable molecular-level insights for the rational design of DESs for targeted applications.

## INTRODUCTION

Deep eutectic solvents (DESs) are an emerging class of sustainable solvents that have gained attention as environmentally friendly alternatives to conventional room-temperature ionic liquids (RTILs)[1]. Unlike RTILs, which consist solely of ions, DESs are typically formed by combining a quaternary ammonium salt with a molecular hydrogen bond donor (HBD), resulting in a eutectic mixture with a melting point significantly lower than that of the individual components.[1] Among the various classes of DESs, those based on choline chloride (ChCl) have proven particularly attractive due to the low cost, minimal toxicity, and high biodegradability [2, 3]. These properties, have led to their widespread application across diverse fields such as electrochemistry, biocatalysis, materials science, and petrochemical processing[4, 5]. While the bulk physicochemical properties of ChCl-based DESs, have been widely explored, microscopic insights into diffusion dynamics, particularly that of cholinium ions remains limited. Notably, choline also holds biological relevance as an essential nutrient involved in various metabolic functions[6, 7] In this study, we employ Quasielastic neutron scattering technique (QENS) to investigate structure-dynamics relationship of cholinium ions in three DES: Reline (ChCl + urea), Glyceline (ChCl + glycerol) and Ethaline (ChCl + ethylene glycol). By systematically varying the hydrogen bond donor (HBD), we aim to elucidate how the nature of the HBD modulates the microscopic diffusion landscape of cholinium ions.

## METHODS AND MATERIALS

DES was synthesized by combining mixtures of choline chloride with deuterated HBDs (urea, glycerol and ethylene glycol) in a molar ratio of 1:2. The mixture was heated to 340K until a clear, homogeneous solution was achieved. Upon cooling to room temperature (300K), the resulting DES remained in liquid state. Quasielastic neutron scattering (QENS) experiments were performed on the IRIS spectrometer at the ISIS Neutron and Muon Source, Rutherford Appleton Laboratory (UK). The IRIS spectrometer utilizes a PG (002) analyzer in offset mode, offering an energy transfer range from -0.3 to +1 meV with an energy resolution of approximately 17 µeV. The accessible wave-vector (Q) transfer range between 0.54 to 1.8 Å$^{-1}$. QENS measurements were carried out at five temperatures - 300, 315, 330, 355 and 365 K. The spectrometer's resolution was calibrated using a standard vanadium sample.

# RESULTS AND DISCUSSION

To elucidate the underlying diffusion mechanism, the QENS spectra was explicitly modelled using a microscopic diffusion mechanism that involves contributions from different dynamic processes. Molecular diffusion in acetamide based DESs has been shown to follow the cage-jump diffusion mechanism, [8, 9,10] wherein acetamide molecules undergo localized motion within transient hydrogen-bonded cages followed by long-range translational jumps. Accordingly, the QENS spectra were modelled using a two-component diffusion model[8] incorporating both localised translation motion within transient cages and jump diffusion of the molecular center of mass. The theoretical scattering function describing these processes is a convolution of their independent scattering functions. However, to fit it with the experimental spectra, the theoretical model function has to be convolution with the instrumental resolution. Therefore, the scattering function used to fit the quasielastic spectra of these DESs can be expressed as

$$S(Q,E) = [A_0(Q)L_j(\Gamma_j, E) + (1 - A_0(Q))L_{j+loc}(\Gamma_j + \Gamma_{loc}, E)] \otimes R(E) \quad (1)$$

where $L_j$ and $L_{j+loc}$ are Lorentzian associated to jump diffusion of cholinium ions (Ch$^+$) COM and cumulative Lorentzian of jump diffusion of Ch$^+$ COM and localised translation diffusion of hydrogen atoms. The fitting parameters obtained from the QENS spectra i.e. $\Gamma_j$, $\Gamma_{j+loc}$ represents the half-width half maxima (HWHM) of the two Lorentzians and $A_0(Q)$ the elastic incoherent structure factor possesses information of the geometry of localised motion. Typical QENS spectra illustrating the influence of HBDs on the microscopic dynamics for all three DESs along with the fits at 315 K at a representative Q is shown in Fig 1. In all systems, the QENS signal predominantly reflects the dynamics of Ch$^+$ due to the high incoherent cross section of hydrogen atoms.

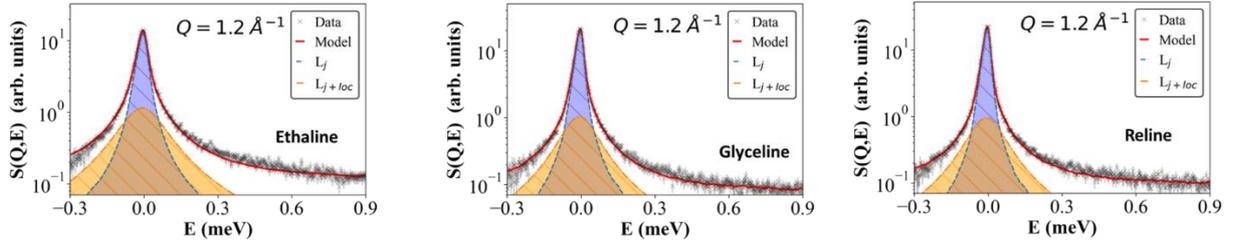

**FIGURE 1.** QENS spectra of ethaline, glyceline and reline at 315 K. The individual components of the model are sown by the shaded regions.

To gain quantitative insight into the mobility of Cl$^+$, the width of the Lorentzian component was modelled using SS using Singwi-Sjolander (SS) model[11] wherein the width of the Lorentzian, $\Gamma_j$ is given by

$$\Gamma_j = \frac{D_j Q^2}{1 + D_j Q^2 \tau} \quad (2)$$

where $D_j$ is the jump diffusion coefficient and $\tau$ is the average time interval between subsequent jumps (also referred to as residence time). The quasielastic widths, $\Gamma_j$, obtained from least square fitting of the QENS spectra with eq. (1) are shown in Fig. 2(a) at 315 K. The solid lines in the figure indicate the fits based on SS model using eq. (2). The obtained jump diffusion coefficient as a function of temperature for all the systems are shown in Fig 2(b). The jump diffusion coefficient and mean residence times obtained for all three DESs using these fits are tabulated in Table 1. Among the three systems, ethaline consistently exhibited the highest jump diffusion coefficient of Ch$^+$ across all temperatures, attributed to its shorter residence times facilitating frequent molecular displacements. In contrast, reline displayed significantly longer residence times than glyceline, which should lead to slower mobility of Ch$^+$ in reline than in glyceline. However, the jump diffusion coefficient of cholinium ions is reline and glyceline were found to be comparable at lower temperatures, and notably, reline surpassed glyceline above 330 K.

To elucidate this behaviour, we calculated the jump length ($l_o$) and found it to be the largest in reline. The jump lengths remained nearly temperature-independent, with average values of 1.05 Å for ethaline, 1.15 Å for glyceline and 1.72 Å for reline. At lower temperatures, the longer residence time in reline limits mobility, but this is offset by its longer jump length, resulting in diffusivity comparable to glyceline. However, with increasing temperature, the enhanced jump frequency, when combined with the larger jump length, leads to a marked increase in diffusivity in reline, ultimately exceeding that in glyceline at elevated temperatures. These findings underscore the crucial role of the interplay between residence time and jump length in governing the microscopic dynamics of DESs, providing valuable insight for tailoring their transport properties through molecular design.

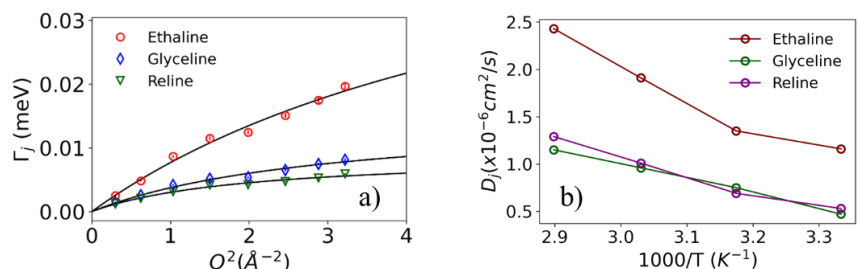

**FIGURE 2.** a) Quasielastic width ($\Gamma_j$) of the Lorentzian ascribed to jump diffusion for all DESs obtained from fitting at 315 K using eq. (2) b) Temperature dependent variation of jump diffusion coefficient for all DESs

**Table 1**. Jump diffusion constant and mean residence time obtained from SS model for all three DESs.

| Temperature (K) | Ethaline | | Glyceline | | Reline | |
|---|---|---|---|---|---|---|
| | $D_j$ (x $10^{-6}$ cm$^2$/s) | $\tau$ (ps) | $D_j$ (x $10^{-6}$ cm$^2$/s) | $\tau$ (ps) | $D_j$ (x $10^{-6}$ cm$^2$/s) | $\tau$ (ps) |
| 300 | 1.16 (±0.09) | 26.43 | 0.47 (±0.05) | 58.07 | 0.53 (±0.06) | 124.02 |
| 315 | 1.35 (±0.12) | 11.86 | 0.75 (±0.08) | 42.90 | 0.69 (±0.07) | 72.78 |
| 330 | 1.91 (±0.12) | 6.88 | 0.96 (±0.12) | 20.07 | 1.01 (±0.13) | 45.75 |
| 345 | 2.43 (±0.19) | 5.32 | 1.15 (±0.13) | 10.12 | 1.29 (±0.16) | 27.39 |

## CONCLUSIONS

The present study underscores the pivotal role of hydrogen bond donor identity in modulating microscopic diffusion dynamics of choline chloride-based deep eutectic solvents. QENS results reveal that the variations in HBD structure critically influences the extent of mobility in DESs influencing the interplay between jump length and residence time of cholinium ions, thereby dictating their overall diffusivity. Ethaline, with its shorter residence times, supports enhanced ion mobility, whereas reline exhibits a temperature-driven crossover due to its longer jump lengths. These findings reveal that subtle variations in HBD structure can lead to pronounced differences in ion transport, emphasizing the need to consider microscopic diffusion mechanisms when tailoring DESs for specific functional environments.